\documentclass[aps,10pt, twocolumn,nofootinbib,superscriptaddress,preprintnumbers]{revtex4}
\usepackage{slashed}
\usepackage{times}
\usepackage{epsfig}
\usepackage{amsmath}
\usepackage{amsfonts}
\usepackage{amssymb}
\newcommand{\be}{\begin{equation}}
\newcommand{\ee}{\end{equation}}
\newcommand{\bea}{\begin{eqnarray}}
\newcommand{\eea}{\end{eqnarray}}

\begin{document}
\preprint{KEK-TH-1164}
\pagestyle{plain}
\title{Unparticle physics and Higgs phenomenology}
\author{Tatsuru Kikuchi}
\affiliation{Theory Division, KEK,
1-1 Oho, Tsukuba, 305-0801, Japan.}
\author{Nobuchika Okada}
\affiliation{Theory Division, KEK,
1-1 Oho, Tsukuba, 305-0801, Japan.}
\affiliation{Department of Physics, University of Maryland, 
College Park, MD 20742, USA
}
\begin{abstract}
Recently, conceptually new physics beyond the Standard Model 
 has been proposed, where a hidden  conformal sector provides 
 ``unparticle'' which couples to the Standard Model sector 
 through higher dimensional operators in low energy effective theory. 
Among several possibilities, we focus on operators 
 involving unparticle, the Higgs doublet and the gauge bosons. 
Once the Higgs doublet develops the vacuum expectation value, 
 the conformal symmetry is broken and as a result, 
 the mixing between unparticle and Higgs boson emerges.
We find that this mixing can cause sizable shifts for the couplings 
 between Higgs boson and a pair of gluons and photons, 
 because these couplings exist only at the loop level 
 in the Standard Model. 
These Higgs couplings are the most important ones 
 for the Higgs boson search at the CERN Large Hadron Collider, 
 and the unparticle physics effects may be observed 
 together with the discovery of Higgs boson. 
\end{abstract}
\maketitle
In spite of the success of the Standard Model (SM) 
 in describing all the existing experimental data, 
 the Higgs boson, which is responsible for the electroweak 
 symmetry breaking, has not yet been directly observed, 
 and is one of the main targets 
 at the CERN Large Hadron Collider (LHC). 
At the LHC, the main production process of Higgs boson is 
 through gluon fusion, and if Higgs boson is light, say 
 $m_h \lesssim 150$ GeV, the primary discovery mode 
 is through its decay into two photons. 
In the SM, these processes occur only at the loop level 
 and Higgs boson couples with gluons and photons very weakly. 

A certain class of new physics models includes 
 a scalar field which is singlet under the SM gauge group. 
In general, such a scalar field can mix with the Higgs boson 
 and also can directly couple with gluons and photons 
 through higher dimensional operators 
 with a cutoff in effective low energy theory.   
Even if the cutoff scale is very high, say, 100-1000 TeV, 
 the couplings with gluons and photons can be comparable 
 to or even larger than those of the Higgs boson 
 induced only at the loop level in the SM. 
This fact implies that if such a new physics exists, 
 it potentially has an impact on Higgs boson phenomenology 
 at the LHC. 
In other words, such a new physics may be observed together 
 with the discovery of Higgs boson.

As one of such models, in this letter, 
 we investigate a new physics recently proposed 
 by Georgi \cite{Georgi:2007ek}, which is described in terms 
 of "unparticle" provided by a hidden conformal sector 
 in low energy effective theory. 
A concrete example of unparticle staff was proposed 
 by Banks-Zaks \cite{Banks:1981nn} many years ago, 
 where providing a suitable number of massless fermions, 
 theory reaches a non-trivial infrared fixed point 
 and a conformal theory can be realized at a low energy. 
Various phenomenological considerations on the unparticle physics 
 have been developed in the literature \cite{U-pheno, U-Higgs, U-Higgs2}.

Basic structure of the unparticle physics is as follows. 
First, we introduce a coupling between the new physics operator 
 ($\cal{O}_{\rm UV}$) with dimension $d_{\rm UV}$ 
 and the Standard Model one (${\cal O}_{\rm SM}$) with dimension $n$, 
\bea
 {\cal L} = \frac{c_n}{M^{d_{\rm UV}+n-4}} 
     \cal{O}_{\rm UV} {\cal O}_{\rm SM} ,  
\eea
where $c_n$ is a dimension-less constant, and $M$ is the energy scale 
 characterizing the new physics. 
This new physics sector is assumed to become conformal 
 at a energy $\Lambda_{\cal U}$, and 
 the operator $\cal{O}_{\rm UV}$ flows to the unparticle operator 
 ${\cal U}$ with dimension $d_{\cal U}$. 
In low energy effective theory, we have the operator of the form, 
\bea
{\cal L}=c_n 
 \frac{\Lambda_{\cal U}^{d_{\rm UV} - d_{\cal U}}}{M^{d_{\rm UV}+n-4}}   
 {\cal U} {\cal O}_{\rm SM} 
\equiv 
  \frac{\lambda_n}{\Lambda^{d_{\cal U}+ n -4}}  {\cal U} {\cal O}_{\rm SM},  
\eea 
where the dimension of the unparticle ${\cal U}$ have been 
 matched by $\Lambda_{\cal U}$ which is induced 
 the dimensional transmutation, 
 $\lambda_n$ is an order one coupling constant
 and $\Lambda$ is the (effective) cutoff scale of 
 low energy effective theory. 
In this paper, we consider only the scalar unparticle. 
It was found in Ref.~\cite{Georgi:2007ek}
 that, by exploiting scale invariance of the unparticle, 
 the phase space for an unparticle operator 
 with the scale dimension $d_{\cal U}$ and momentum $p$ 
 is the same as the phase space for 
 $d_{\cal U}$ invisible massless particles, 
\begin{eqnarray}
d \Phi_{\cal U}(p) = 
 A_{d_{\cal U}} \theta(p^0) \theta(p^2)(p^2)^{d_{\cal U}-2} 
 \frac{d^4p}{(2\pi)^4}\,,
\end{eqnarray}
where
\begin{eqnarray}
A_{d_{\cal U}} = \frac{16 \pi^{\frac{5}{2}}}{(2\pi)^{2 d_{\cal U}}}
\frac{\Gamma(d_{\cal U}+\frac{1}{2})}{\Gamma(d_{\cal U}-1) 
\Gamma(2 d_{\cal U})}.
\end{eqnarray}
%
Also, based on the argument on the scale invariance, 
 the (scalar) propagator for the unparticle was suggested to be 
\begin{eqnarray}
 \frac{A_{d_{\cal U}}}{2\sin(\pi d_{\cal U})}
 \frac{i}{(p^2)^{2-d_{\cal U}}} 
 e^{-i (d_{\cal U}-2) \pi}  \,.
\label{propagator}
\end{eqnarray}
%
Interestingly, $d_{\cal U}$ is not necessarily integral, 
 it can be any real number or even complex number. 
In this paper we consider the scale dimension 
 in the range,  $1 <  d_{\cal U} <  2$, 
 for simplicity. 

Among several possibilities, we focus on the operators 
 between the unparticle and the Higgs sector, 
\bea 
 {\cal L} = \frac{\lambda_n}{\Lambda^{d_{\cal U}+ n - 4}} 
 {\cal U} {\cal O}_{\rm SM}(H^\dagger H), 
\eea 
where $H$ is the Standard Model Higgs doublet and  
 ${\cal O}_{\rm SM}(H^\dagger H)$ is the Standard Model 
 operator as a function of the gauge invariant 
 bi-linear of the Higgs doublet. 
Once the Higgs doublet develops the vacuum expectation value (VEV), 
 the tadpole term for the unparticle operator is induced, 
\bea 
{\cal L}_{\slashed{\cal U}} =
 \Lambda_{\slashed{\cal U}}^{4-d_{{\cal U}}} {\cal U},  
\label{tadpole} 
\eea 
 and the conformal symmetry in the new physics sector is broken 
 \cite{U-Higgs}. 
Here, 
   $ \Lambda_{\slashed{\cal U}}^{4-d_{{\cal U}}}= \lambda_n
   \langle {\cal O}_{\rm SM} \rangle/ \Lambda^{d_{\cal U}+n-4}$ 
 is the conformal symmetry breaking scale. 
At the same time, we have the interaction terms 
 between the unparticle and the physical Standard Model Higgs boson 
 ($h$) such as (up to ${\cal O}(1)$ coefficients) 
\bea 
  {\cal L}_{{\cal U}-{\rm Higgs}} 
  = \left(\Lambda_{\slashed {\cal U}}^{4-d_{{\cal U}}}/v \right) 
   {\cal U} h 
  + \left(\Lambda_{\slashed {\cal U}}^{4-d_{{\cal U}}}/v^2 \right)
    {\cal U} h^2 + \cdots,   
 \label{mixing}
\eea 
where $v=246$ GeV is the Higgs VEV. 
In order not to cause a drastic change or instability in the Higgs potential, 
 the scale of the conformal symmetry breaking  
 should naturally be smaller than the Higgs VEV, 
 $\Lambda_{\slashed {\cal U}} \lesssim v$.

As other operators between the unparticle and the Standard Model sector, 
 we consider 
\bea
{\cal L}_{\cal U} = 
 -\frac{\lambda_g}{4}  \frac{\cal U} {\Lambda^{d_{\cal U}}}
      G^A_{\mu \nu} G^{A \mu \nu} 
 -\frac{\lambda_\gamma}{4} \frac{\cal U}{\Lambda^{d_{\cal U}}} 
      F_{\mu \nu} F^{\mu \nu},  
\label{Unp-gauge} 
\eea
where we have neglected ${\cal O}$(1) coefficients, but 
 taken into account of the two possible relative signs 
 of the coefficients, 
 $\lambda_g = \pm 1$ and $\lambda_\gamma= \pm 1$. 
We will see that these operators are the most important ones 
 relevant to the Higgs phenomenology. 

Now let us focus on effective couplings 
 between the Higgs boson and the gauge bosons 
 (gluons and photons) of the form, 
\bea 
 {\cal L}_{\rm Higgs-gauge} = 
  \frac{1}{v} C_{gg} \;  h G^A_{\mu \nu} G^{A \mu \nu}  
+ \frac{1}{v} C_{\gamma \gamma} \; h F_{\mu \nu} F^{\mu \nu}.   
 \label{Higgs-gauge}
\eea
As is well-known, in the Standard Model, these operators are induced 
 through loop corrections in which fermions and W-boson are running 
 \cite{HHG}. 
For the coupling between the Higgs boson and gluons, 
 the contribution from top quark loop dominates and is 
 described as\footnote{
In our numerical analysis, we take all fermion contributions 
 into account, for completeness.}
\bea 
 C_{gg}^{\rm SM} = \frac{\alpha_s}{16 \pi} F_{1/2}(\tau_t), 
 \label{SM-glue}
\eea
where $\alpha_s$ is the QCD coupling, 
 and $\tau_t = 4 m_t^2/m_h^2$ with the top quark mass $m_t$ 
 and the Higgs boson mass $m_h$.  
For the coupling between the Higgs boson and photons, 
 there are two dominant contributions from 
loop corrections through top quark and W-boson, 
\bea 
 C_{\gamma \gamma}^{\rm SM} = 
  \frac{\alpha} {8 \pi} 
  \left( 
   \frac{4}{3} F_{1/2}(\tau_t) + F_{1}(\tau_W) 
  \right) ,     
\label{SM-gamma}
\eea 
where $\tau_W = 4 M_W^2/m_h^2$ with the W-boson mass $M_W$. 
In these expressions, the structure functions are given by  
\begin{eqnarray}
F_{1/2}(\tau) &=& 
  2 \tau \left[ 1+ \left( 1 - \tau  \right) f(\tau) \right] , 
 \nonumber\\
F_{1}(\tau) &=& 
 -\left[2 + 3 \tau + 3 \left( 2 - \tau  \right) f(\tau) \right] 
\end{eqnarray}
with 
\begin{eqnarray}
f(\tau)  =  
   \left\{ 
\begin{array}{cc}
\left[ \sin^{-1}\left( 1/\sqrt{\tau}\right)\right]^2 &    ({\rm for}~\tau\ge 1), \\

-\frac{1}{4} \left[ \ln \left( \frac{1+\sqrt{1-\tau}}{1-\sqrt{1-\tau}} \right) - i \pi \right]^2 &
               ({\rm for}~\tau < 1) . 
\end{array}  
\right.   
\nonumber 
\end{eqnarray}
Note that even though the effective couplings are loop suppressed 
 in the Standard Model, they are the most important ones 
 for the Higgs boson search at the LHC. 
In the wide range of the Higgs boson mass $m_h < 1$ TeV, 
 the dominant Higgs boson production process at the LHC 
 is the gluon fusion channel through the first term 
 in Eq.~(\ref{Higgs-gauge}). 
If the Higgs boson is light, $m_h < 2 M_W$, 
 the primary discovery mode of the Higgs boson is 
 on its decay into two photons, 
 in spite of this branching ratio is ${\cal O}(10^{-3})$ at most. 
Therefore, a new physics will have a great impact on 
 the Higgs phenomenology at LHC  
 if it can provide sizable contributions to 
 the effective couplings in Eq.~(\ref{Higgs-gauge}). 
Furthermore, the fact that the Standard Model contributions 
 are loop-suppressed implies that it is relatively easier 
 to obtain sizable (or sometimes big) effects from new physics. 

Now we consider new contributions to the Higgs effective couplings 
 induced through the mixing between the unparticles 
 and the Higgs boson (the first term in Eq.~(\ref{mixing})) 
 and Eq.~(\ref{Unp-gauge}), 
 in other words, through the process  
 $h \to {\cal U} \to gg$  or $\gamma \gamma$. 
We can easily evaluate them 
  by using the vertex among the unparticle, 
 the Higgs boson and gauge bosons and the unparticle propagator as 
\bea
 C^{\cal U}_{gg, \gamma \gamma} 
&=& \Lambda_{\slashed{\cal U}}^{4-d_{\cal U}} 
   \times
\left(
   \frac{A_{d_{\cal U}}}{2\sin(\pi d_{\cal U})} 
   \frac{e^{-i (d_{\cal U}-2) \pi}}{(m_h^2)^{2-d_{\cal U}}}  \right)  \times
\left( 
  \frac{\lambda_{g, \gamma}}{\Lambda^{d_{\cal U}}}
  \right)                                 
\nonumber\\
 &=&   \lambda_{g, \gamma} \; 
 \frac{A_{d_{\cal U}} e^{-i (d_{\cal U}-2) \pi}}{2\sin(\pi d_{\cal U})}
  \left(\frac{\Lambda_{\slashed {\cal U}}}{m_h} \right)^{4-d_{{\cal U}}}
  \left(\frac{m_h}{\Lambda} \right)^{d_{\cal U}} \;,  \quad \quad
\label{Unp-g}
\eea
where we replaced the momentum in the unparticle propagator 
 into the Higgs mass, $p^2 = m_h^2$. 
The unparticle contributions become smaller 
 as $m_h$ and $\Lambda$ ($\Lambda_{\slashed{\cal U}}$) 
 become larger (smaller) 
 for a fixed $1< d_{\cal U} <2$. 
Note that in the limit $d_{\cal U} \to 1$, 
 the unparticle behaves as a real scalar field and 
 the above formula reduces into the one obtained 
 through the mass-squared mixing $ \Lambda_{\slashed {\cal U}}^3/v$  
 between the real scalar and the Higgs boson.

Let us first show the partial decay width 
 of the Higgs boson into two gluons and two photons. 
Here we consider the ratio of the sum of the Standard Model 
 and unparticle contributions to the Standard Model one, 
\bea 
 R= \frac{ \Gamma^{{\rm SM}+{\cal U}}(h \to gg,\; \gamma \gamma)} 
      { \Gamma^{\rm SM}(h \to gg,\; \gamma \gamma)} 
 = \frac{
   |C^{\rm SM}_{gg,\; \gamma \gamma} +C^{\cal U}_{gg,\; \gamma \gamma}|^2 
   } 
   {|C^{\rm SM}_{gg,\; \gamma \gamma}|^2} . 
\eea
Using Eqs.~(\ref{SM-glue}), ({\ref{SM-gamma}) and (\ref{Unp-g}) 
 we evaluate the ratio of the partial decay widths 
 as a function of $d_{\cal U}$. 
The results are shown in Fig.~1 and Fig.~2, 
 for $\Lambda_{\slashed{\cal U}}=v$ and $m_h$=120 GeV, 
 and different choices of $\Lambda = 100$ and $1000$ TeV. 
Even for $\Lambda={\cal O}$(1000 TeV), 
 we can see a sizable deviation of ${\cal O}(10 \%)$ 
 from the Standard Model one with $d_{\cal U} \sim 1$. 
Here, it is shown that the relative sign $\lambda_{g, \gamma}$ play 
 an important role in the interference between 
 the unparticle and the Standard Model contributions. 

\begin{figure}[h]
\includegraphics[width=0.8\columnwidth]{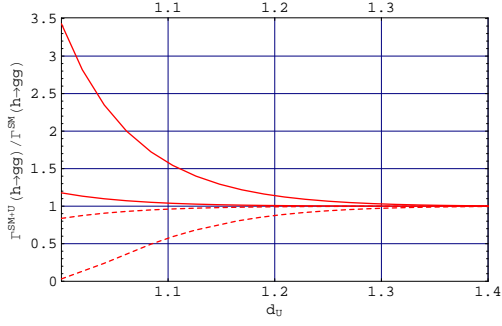}
\caption{
The ratio of the partial decay width for the mode $h \to gg$ 
 as a function of $d_{\cal U}$ for $\Lambda = 100$ and $1000$ TeV, 
 with $\Lambda_{\slashed{\cal U}}=v$ and $m_h=120$ GeV. 
The solid and dashed lines correspond to 
 $\lambda_g= +$ and $\lambda_g= -$, respectively. 
Each solid or dashed line with larger deviations 
 corresponds to $\Lambda = 100$ TeV. 
}
\end{figure}
%
\begin{figure}[h]
\includegraphics[width=0.8\columnwidth]{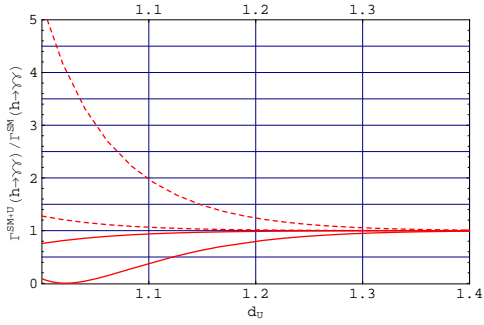}
\caption{
The ratio of the partial decay width 
 for the mode $h \to \gamma \gamma$ 
 as a function of $d_{\cal U}$ for $\Lambda = 100$ and $1000$ TeV. 
The solid line and dashed line correspond to 
 $\lambda_\gamma= +$ and $\lambda_\gamma= -$, respectively. 
Each solid or dashed line with larger deviations 
 corresponds to $\Lambda = 100$ TeV. 
}
\end{figure}

Assuming $m_h < 2 M_W$, 
 it is interesting to evaluate the Higgs boson signal events 
 through $ gg \to h \to \gamma \gamma$ at the LHC, 
 in the presence of the unparticle effects. 
We consider the ratio of two photon events 
 including unparticle effects to those in the Standard Model. 
For $\Lambda \gtrsim 100$ TeV,  
 we will approximately evaluate the event number ratio. 
Note that the branching ratio into 
 $h \to gg, \; \gamma \gamma$ is small in the Standard Model 
 and the unparticle contributions are comparable to 
 or smaller than the Standard Model ones (see Fig.~1 and 2). 
Thus, the deviation of the total decay width 
 due to the unparticle effects is negligible, 
 and we arrive at the approximation formula 
 for the event number ratio ($r$), 
\bea 
 r &=& 
 \frac{
 \sigma^{{\rm SM}+{\cal U}}(gg \to h)  \times 
 {\rm BR}^{{\rm SM}+{\cal U}}(h \to \gamma \gamma) 
 } {
 \sigma^{\rm SM}(gg \to h)  \times 
 {\rm BR}^{\rm SM}(h \to \gamma \gamma) } \nonumber  \\
&\simeq & 
 \frac{ 
 \Gamma^{{\rm SM}+{\cal U}} (h \to gg)  \times 
 \Gamma^{{\rm SM}+{\cal U}} (h \to \gamma \gamma) 
 } {
 \Gamma^{\rm SM} (h \to gg)  \times 
 \Gamma^{\rm SM} (h \to \gamma \gamma) } \nonumber \\ 
&=&
 \frac{ 
 |C^{\rm SM}_{gg}+C^{\cal U}_{gg}|^2 \times 
 |C^{\rm SM}_{\gamma \gamma} + C^{\cal U}_{\gamma \gamma}|^2 
 }{ 
 |C^{\rm SM}_{gg}|^2  \times |C^{\rm SM}_{\gamma \gamma}|^2 
 } .
\label{EventRatio}
\eea
The event number ratio as a function of $d_{{\cal U}}$ 
 are depicted in Fig.~3 and 4 
 for $\Lambda=100$ and $1000$ TeV, respectively, 
 with $\Lambda_{\slashed{\cal U}}=v$ and $m_h=120$ GeV. 
\begin{figure}[h]
\includegraphics[width=0.8\columnwidth]{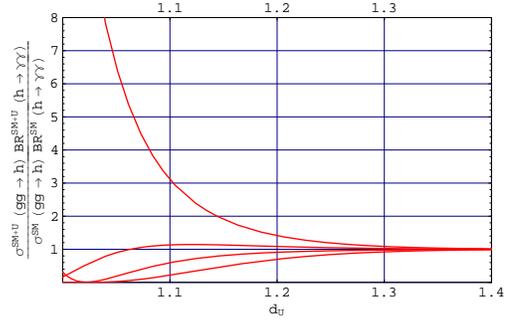}
\caption{
The ratio of the two photon evens as a function of $d_{{\cal U}}$
 for $\Lambda = 100$ TeV, 
 with $\Lambda_{\slashed{\cal U}}=v$ and $m_h=120$ GeV. 
Each line from top to bottom at $d_{\cal U}=1.1$ 
 corresponds to 
 $(\lambda_g,~\lambda_\gamma)=(+,-)$, $(-,-)$, $(+,+)$ and 
 $(-,+)$, respectively. 
}
\end{figure}
%
\begin{figure}[h]
\includegraphics[width=0.8\columnwidth]{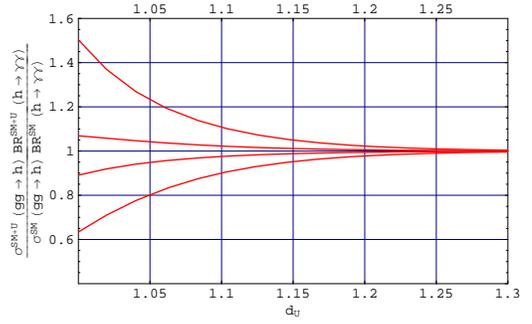}
\caption{
The same figure as Fig.~3 
 but for $\Lambda = 1000$ TeV. 
}
\end{figure}

As discussed before, once the Higgs doublet develops the VEV, 
 the conformal symmetry is broken in the new physics sector, 
 providing the tadpole term in Eq.~(\ref{tadpole}). 
Once such a tadpole term is induced, 
 the unparticle will subsequently develop the VEV
 \cite{U-Higgs, U-Higgs2} 
 whose order is naturally the same as the scale of 
 the conformal symmetry breaking, 
\bea
  \langle {\cal U} \rangle 
= \left( c \; \Lambda_{\slashed {\cal U}} \right)^{d_{{\cal U}}} . 
\eea 
Here we have introduced a numerical factor $c$, 
 which can be $c = {\cal O}(0.1) - {\cal O}(1)$, 
 depending on the naturalness criteria. 
Through this conformal symmetry breaking, 
 parameters in the model are severely constrained 
 by the current precision measurements. 
We follow the discussion in Ref.~\cite{U-Higgs2}. 
 From Eq.~(\ref{Unp-gauge}), 
 the VEV of the unparticle leads to the modification of 
 the photon kinetic term, 
\bea
{\cal L} =
-\frac{1}{4}\left[ 
 1 \pm \frac{\langle {\cal U} \rangle}{\Lambda^{d_{{\cal U}}}}
\right] F_{\mu \nu} F^{\mu \nu} ,
\eea
 which can be interpreted as a threshold correction 
 in the gauge coupling evolution across the scale 
 $\langle {\cal U} \rangle^{1/d_{\cal U}}$. 
The evolution of the fine structure constant from 
 zero energy to the Z-pole is consistent 
 with the Standard Model prediction, 
 and the largest uncertainty arises 
 from the fine structure constant 
 measured at the Z-pole \cite{Yao:2006px}, 
\bea
 \widehat{\alpha}^{-1}(M_Z) &=&  127.918 \pm 0.019 . 
\nonumber 
\eea
This uncertainty (in the $\overline{\rm MS}$ scheme) 
 can be converted to the constraint,  
\bea
\epsilon=
\left<{\cal U} \right>/\Lambda^{d_{{\cal U}}}
 \lesssim 1.4 \times 10^{-4}. 
\label{gauge}
\eea
This provides a lower bound on the effective cutoff scale. 
For $d_{\cal U} \simeq 1$ and $\Lambda_{\slashed{\cal U}} \simeq v$ 
 we find 
\bea
\Lambda \gtrsim c \times 1000~{\rm TeV} ,  
 \nonumber 
\eea
This is a very severe constraint on the scale of new physics, 
 for example, $\Lambda \gtrsim 100$ TeV for $c \gtrsim 0.1$.

A similar bound can be obtained by the results on  
 Higgs boson search through two photon decay mode 
 at the Tevatron. 
With the integrated luminosity 1 fb$^{-1}$ and 
 the Higgs boson mass around $m_h=120$ GeV for example,  
 the ratio in Eq.~(\ref{EventRatio}) is constrained to be 
 $r \lesssim 50$ \cite{Wells}. 
For $d_{\cal U} \simeq 1$, this leads to the bound, 
 $\Lambda \gtrsim 60$ TeV, which is, as far as we know, 
 the strongest constraint on the cutoff scale 
 by the current collider experiments.

In conclusion, we have considered the unparticle physics 
 focusing on the Higgs phenomenology. 
Once the electroweak symmetry breaking occurs, 
 the conformal symmetry is also broken and 
 this breaking leads to the mixing between the unparticle 
 and the Higgs boson. 
Providing the operators among the unparticle and the gauge bosons 
 (gluons and photons),  
 the unparticle brings the sizable deviation into 
 effective couplings between the Higgs boson and the gauge bosons, 
 that can be measured at the LHC through the discovery of the Higgs boson. 
The conformal symmetry breaking induces threshold corrections 
 for the gauge coupling evolutions, 
 and the current precision measurements on the fine structure constant 
 require the effective cutoff scale far above the electroweak scale. 
The similar bound on the cutoff scale can be obtained 
 from the Tevatron experiments. 
When we naively assume the common cutoff scale for operators 
 between the unparticle and the Standard Model sector,  
 it seems very hard to measure unparticle effects in any processes. 
However, as have been shown in this paper, 
 there exist sizable effects in the Higgs phenomenology. 
This is the point of this paper. 
The unparticle physics makes Higgs phenomenology more interesting.

Finally, we give several comments. 
In Eq.~(\ref{Unp-gauge}), we can in general introduce 
 the coupling involving the weak gauge bosons 
 and also discuss the unparticle effects on couplings 
 between the Higgs bosons and the weak gauge bosons. 
However, such effects are negligible, because the Higgs boson 
 couples to the weak gauge boson at tree level and 
 the effective cutoff is required to be very high 
 by the current experiments. 
In this sense, we can neglect several operators 
 through which the unparticle cause additional contributions 
 to the couplings existing in the Standard Model at tree level. 

If there exists a coupling between the unparticle 
 and Yukawa sectors in the Standard Model, 
 the unparticle VEV gives an additional contribution 
 to fermion masses. 
In order for such a fermion mass contribution 
 to be consistent with the observed fermion masses, 
 the coefficient of the coupling involving light fermions 
 should be very small. 
As a result, the Yukawa coupling constant of the interactions term 
 ${\cal U}\bar{\Psi}_f \Psi_f$ 
 is at most $m_f/v$, where $m_f$ is the fermion mass, 
 and negligibly small for light fermions.

There are other new physics models which can cause 
 the similar effects in Higgs phenomenology 
 as what we have investigated in this paper. 
A well-studied example is two Higgs doublet model (2HDM) 
 where the SM Higgs sector is simply modified. 
In fact, there has been a study on a comparison 
 between the 2HDM and the SM results in the effective 
 Higgs coupling to two photons \cite{2HDM}. 
It is worth investigating how to distinguish 
 the unparticle physics from 2HDM. 
A clear difference is that the unparticle 
 is singlet under the SM gauge group, 
 while the charged Higgs bosons appears in 2HDM. 
However, in the part involving only (CP-even) 
 neutral Higgs bosons, 2HDM has many free parameters 
 enough to produce the same results we have obtained, 
 by choosing a special parameter set. 

Once the unparticle develops the VEV, the unparticle 
 may acquire a ``mass'' term characterized by the scale 
 $\Lambda_\slashed{\cal U}$ \cite{U-Higgs}. 
If this is the case, it is interesting to investigate 
 the unparticle production at the LHC. 
Through Eq.~(\ref{Unp-gauge}), it is produced via gluon fusion 
 and subsequently decays into two gluons and photons. 
Therefore, the unparticle behaves like the Higgs boson at the LHC. 
However, counting the degrees of freedom of the gluons 
 and photons, the branching ratio into photon will be  
 $1/9 \sim 0.1$, so that we can expect two photon events 
 from the unparticle production much larger than 
 those from the Higgs boson. 
This unparticle phenomenology is similar to the one 
 investigated in Ref.~\cite{Itoh:2006fv}, 
 where the scalar particle in the supersymmetry breaking sector 
 plays the similar role of the unparticle 
 in the limit $d_{\cal U} \to 1$. 

T.K. would like to thank K.S. Babu and R.N. Mohapatra 
 for their hospitalities at Oklahoma State University and 
 University of Maryland, respectively. 
The work of N.O. is supported in part by 
 the Grant-in-Aid for Scientific Research from the Ministry 
 of Education, Science and Culture of Japan (\#18740170). 
The work of T.K. was supported by the Research Fellowship 
 of the Japan Society for the Promotion of Science (\#1911329).

\end{document}